%% file: ISMIR2025_strumming.tex
\documentclass{article}
\usepackage{subcaption}
\usepackage[T1]{fontenc}
\usepackage[utf8]{inputenc}
\usepackage[]{ismir} 
\usepackage{amsmath,cite,url}
\usepackage{graphicx}
\usepackage{color}

\usepackage{algorithmic}
\usepackage{textcomp}
\usepackage{xcolor}

\usepackage{siunitx}
\usepackage{float}
\usepackage{tikz}
\usetikzlibrary{shapes.geometric, arrows, positioning}
\usepackage{booktabs}
\usepackage{microtype}  

\definecolor{highlight1}{HTML}{0077BB} 
\definecolor{highlight2}{HTML}{EE7733} 
\definecolor{highlight3}{HTML}{009988} 
\definecolor{highlight4}{HTML}{CC3311} 
\definecolor{highlight5}{HTML}{AA4499} 
\definecolor{highlight6}{HTML}{EECC66} 

\title{Joint Transcription of Acoustic Guitar Strumming Directions and Chords}

\multauthor
  {Sebastian Murgul$^{1,2}$ \hspace{1cm} Johannes Schimper$^2$ \hspace{1cm} Michael Heizmann$^2$}
  {
  $^1$ Klangio GmbH, Karlsruhe, Germany\\
  $^2$ Karlsruhe Institute of Technology, Karlsruhe, Germany\\
  {\tt\small sebastian.murgul@klang.io}
  }

\def\authorname{S. Murgul, J. Schimper, and M. Heizmann}

\begin{document}

\maketitle

\begin{abstract}%
  \input{sections/000_abstract.tex}%
\end{abstract}%
\input{sections/001_introduction.tex}
\input{sections/002_multimodal.tex}%
\input{sections/003_synthesis.tex}%
\input{sections/004_model.tex}%
\input{sections/005_experiments.tex}%
\input{sections/006_conclusion.tex}%
\newpage%
%
\bibliography{ISMIR2025_strumming}%
\end{document}

%% file: sections/000_abstract.tex
Automatic transcription of guitar strumming is an underrepresented and challenging task in Music Information Retrieval (MIR), particularly for extracting both strumming directions and chord progressions from audio signals. While existing methods show promise, their effectiveness is often hindered by limited datasets. In this work, we extend a multimodal approach to guitar strumming transcription by introducing a novel dataset and a deep learning-based transcription model. We collect $\SI{90}{\minute}$ of real-world guitar recordings using an ESP32 smartwatch motion sensor and a structured recording protocol, complemented by a synthetic dataset of $\SI{4}{\hour}$ of labeled strumming audio. A Convolutional Recurrent Neural Network (CRNN) model is trained to detect strumming events, classify their direction, and identify the corresponding chords using only microphone audio. Our evaluation demonstrates significant improvements over baseline onset detection algorithms, with a hybrid method combining synthetic and real-world data achieving the highest accuracy for both strumming action detection and chord classification. These results highlight the potential of deep learning for robust guitar strumming transcription and open new avenues for automatic rhythm guitar analysis.

%% file: sections/001_introduction.tex
\section{Introduction}
\label{sec:introduction}
Automatic music transcription is a key task in Music Information Retrieval (MIR), aiming to convert audio signals into symbolic representations.
For the transcription of solo instrument music, numerous new approaches and tools have been proposed over the last years \cite{benetos2019automatic}.
While classical note-tracking models such as \cite{riley2024high}, \cite{chang2024ymt3}, and \cite{wiggins2019guitar} perform well for fingerpicking, they are not designed to predict strumming directions. These models focus on individual note onsets and often struggle with the dense polyphony and rhythmic structure of strumming, where the emphasis lies on chord-level articulation.
This limitation highlights the need for a dedicated strumming transcription system with applications in music education, DAW plugins, and notation software.

Research on guitar strumming transcription has primarily followed two main approaches: audio-based classification and sensor-based motion analysis.
In 2019, Bello et al.\ proposed a neural network-based classification system to distinguish between up and down strokes using Mel-Frequency Cepstral Coefficients (MFCCs) segments as input features \cite{bello2019classification}. Their approach achieved a classification accuracy of $\SI{72.5}{\percent}$ for a Convolutional Neural Network (CNN) and $\SI{70}{\percent}$ for a Long Short-Term Memory (LSTM) model.
Earlier, in 2013, Matsushita et al.\ developed a wristwatch-like device designed to analyze down-strumming actions in terms of note timing and intensity \cite{matsushita2013detecting}. More recently, Freire et al.\ (2020) explored strumming gestures in greater detail using inertial measurement units (IMUs) and motion capture technology, further advancing sensor-based analysis of guitar performance \cite{freire2020evaluation}.
A multimodal approach was introduced in 2022 by Murgul et al., who combined a back-of-hand-mounted motion sensor with guitar pickup audio for strumming action transcription \cite{murgul2022multimodal}. Their method involved recording a small manually labeled dataset, which was used to evaluate algorithmic annotation techniques based on onset detection in the pickup signal and thresholding the first-order derivative of the motion data.

Building on the approach in Murgul et al., we extend the multimodal approach to create a bigger and more diverse dataset in order to train a neural network.
We increase the dataset size from $\SI{5}{\minute}$ to $\SI{90}{\minute}$ and from $4$ chords to $24$ chords (major / minor) while also adding more complex strumming rhythms and performance parameter variations.
Therefore, an improved hand motion sensor based on an off-the-shelf ESP32 smartwatch module is developed, and a sophisticated recording plan with specific instructions to the players is created. A new guitar strumming dataset is recorded by three guitar players using this approach and semi-automatically annotated using the multimodal information.
While the semi-automatic annotation process is scalable, the recording process still does take some time.
Therefore, to complement the real-world dataset, we present a guitar strumming data synthesis approach that is used to generate an additional $\SI{4}{\hour}$ of labeled strumming audio.
These datasets are then used to train a CRNN model to automatically detect strumming events and classify the strumming direction as well as the played chord from solely microphone audio.
Finally, the transcription results are evaluated using the test split of the real strumming recordings and compared with baseline algorithms.

%% file: sections/002_multimodal.tex
\section{Multimodal Strumming Recording}
\label{sec:multimodal}

\subsection{Motion Recording Hardware}

To capture hand movement and, consequently, the strumming direction, a compact and lightweight system is required that can be attached to the playing hand. It must enable wireless communication for transmitting motion data and be capable of starting and stopping audio recordings on a computer via wireless commands. Additionally, the system should be intuitive for guitarists to use. For scalable applications, the solution should be cost-efficient. The ESP32-S3-Touch-LCD-1.28 module from Waveshare meets these requirements and serves as the central microcontroller \cite{waveshare}. It features a 3-axis accelerometer (QMI8658), a LiPo battery connector with a battery management, and supports the wireless standards Wi-Fi and Bluetooth Low Energy (BLE). Furthermore, the module includes an LCD screen with touch functionality and a compact form factor.

\begin{figure}[h]
  \centering
  \begin{subfigure}[c]{0.49\linewidth}
    \centering
    \includegraphics[alt={A compact sensor module designed for hand motion tracking, shown without its protective cover. The small device features a flat rectangular body with visible electronic components.},width=\linewidth]{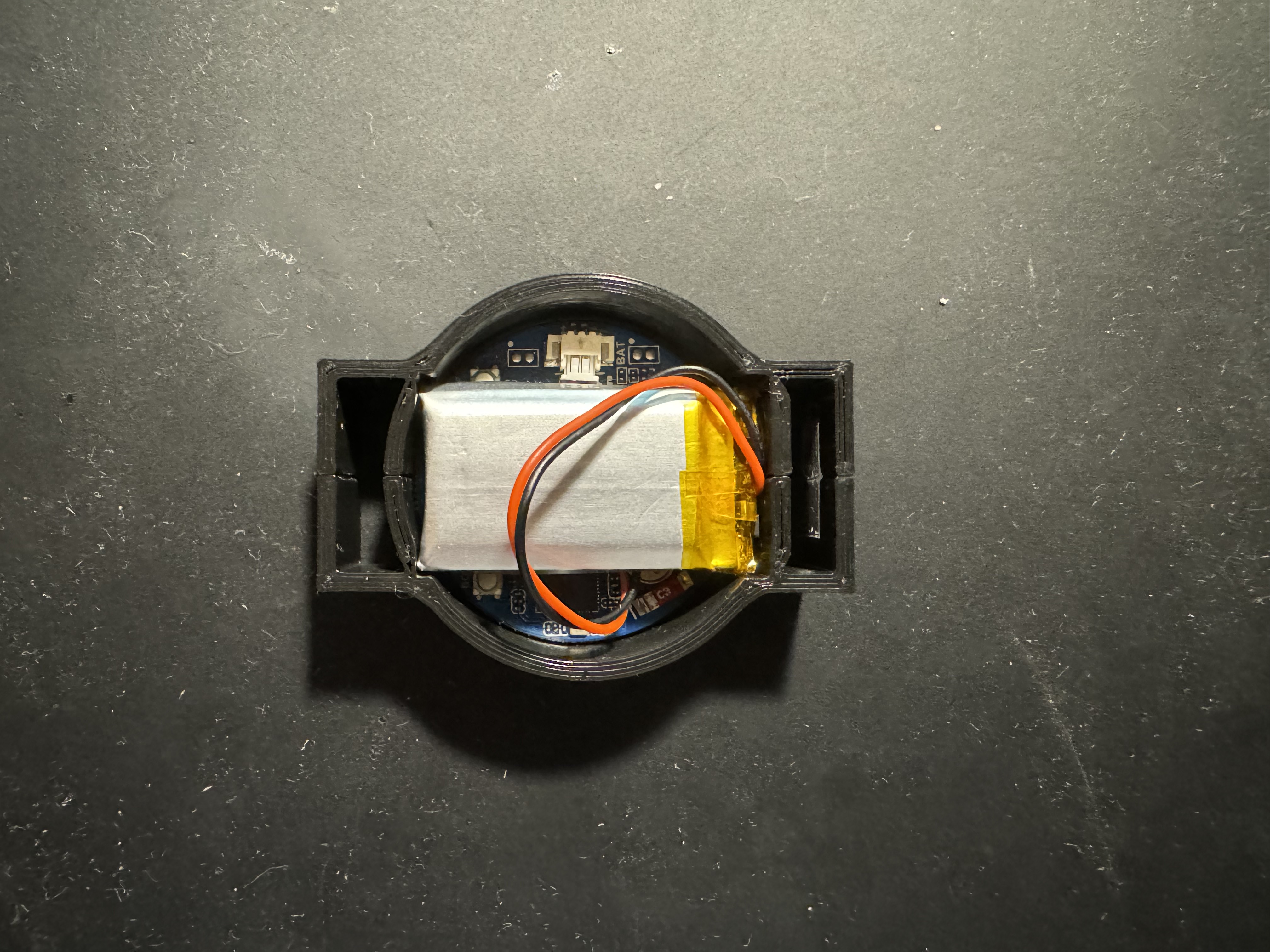}
    \subcaption{Backside of the hand sensor without cover. }
    \label{fig:Gehaeuse2}
  \end{subfigure}
  \hfill
  \begin{subfigure}[c]{0.49\linewidth}
    \centering
    \includegraphics[alt={The hand motion sensor is strapped to the back of a person's hand using a watch-like attachment. The device is secured to ensure stability during guitar playing.},width=\linewidth]{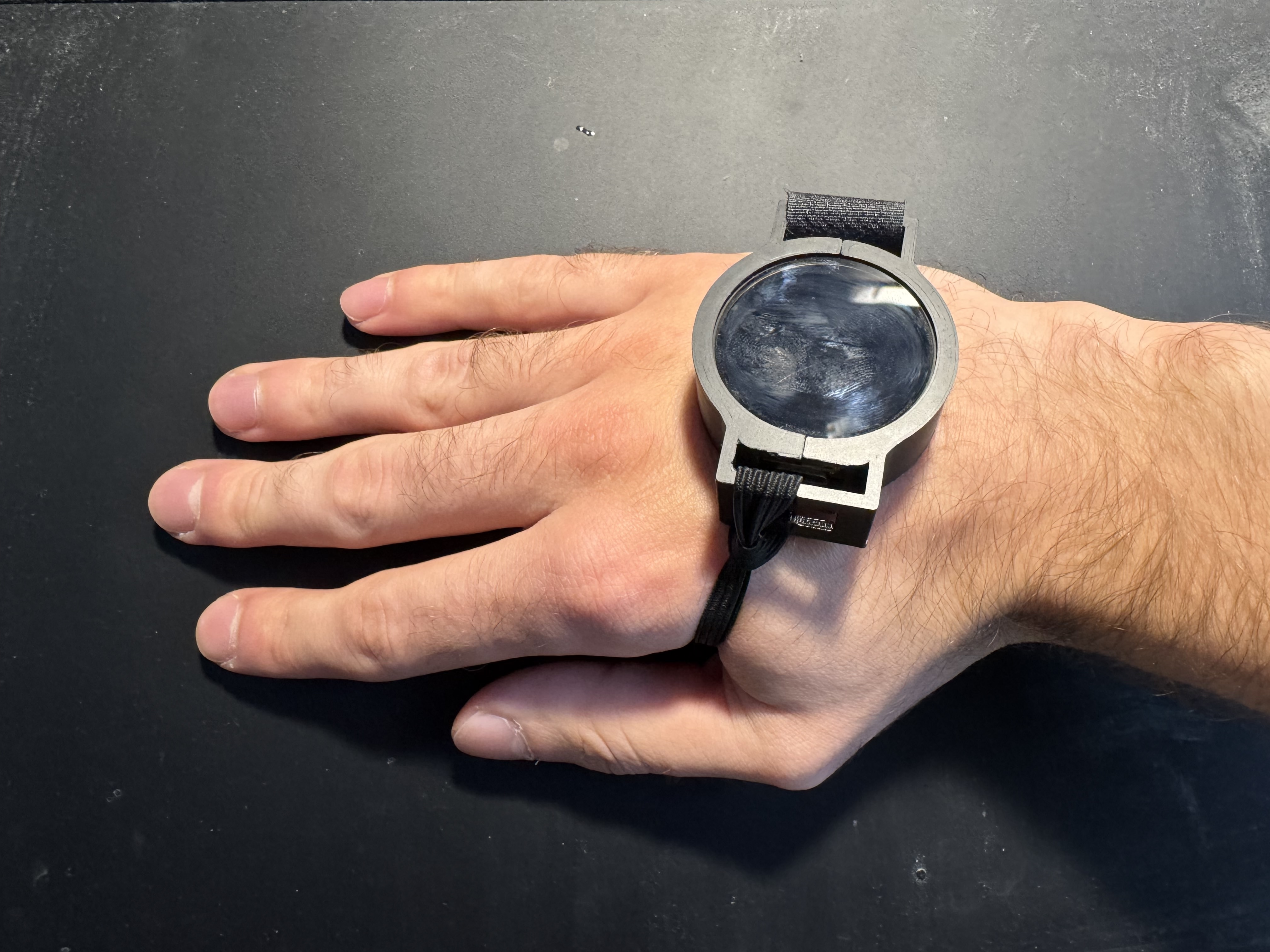}
    \subcaption{Sensor attached to the back of the hand.}
    \label{fig:SensorAnHand}
  \end{subfigure}
  \caption{Hand sensor in its enclosure}
  \label{fig:Sensor}
\end{figure}

A custom 3D-printed enclosure enables a watch-like attachment on the back of the hand. The enclosure also houses a $\SI{350}{\milli\ampere{}\hour}$ LiPo battery, as shown in Figure \ref{fig:Gehaeuse2}. Figure \ref{fig:SensorAnHand} illustrates the sensor system attached to the back of the hand.

\begin{figure}[t]
  \centering
  \includegraphics[alt={A diagram illustrating the motion model of the hand sensor during guitar strumming. The hand's movement follows a semicircular trajectory around the elbow, with labeled axes and forces such as gravitational, centripetal, and tangential acceleration.},width=0.3\textwidth]{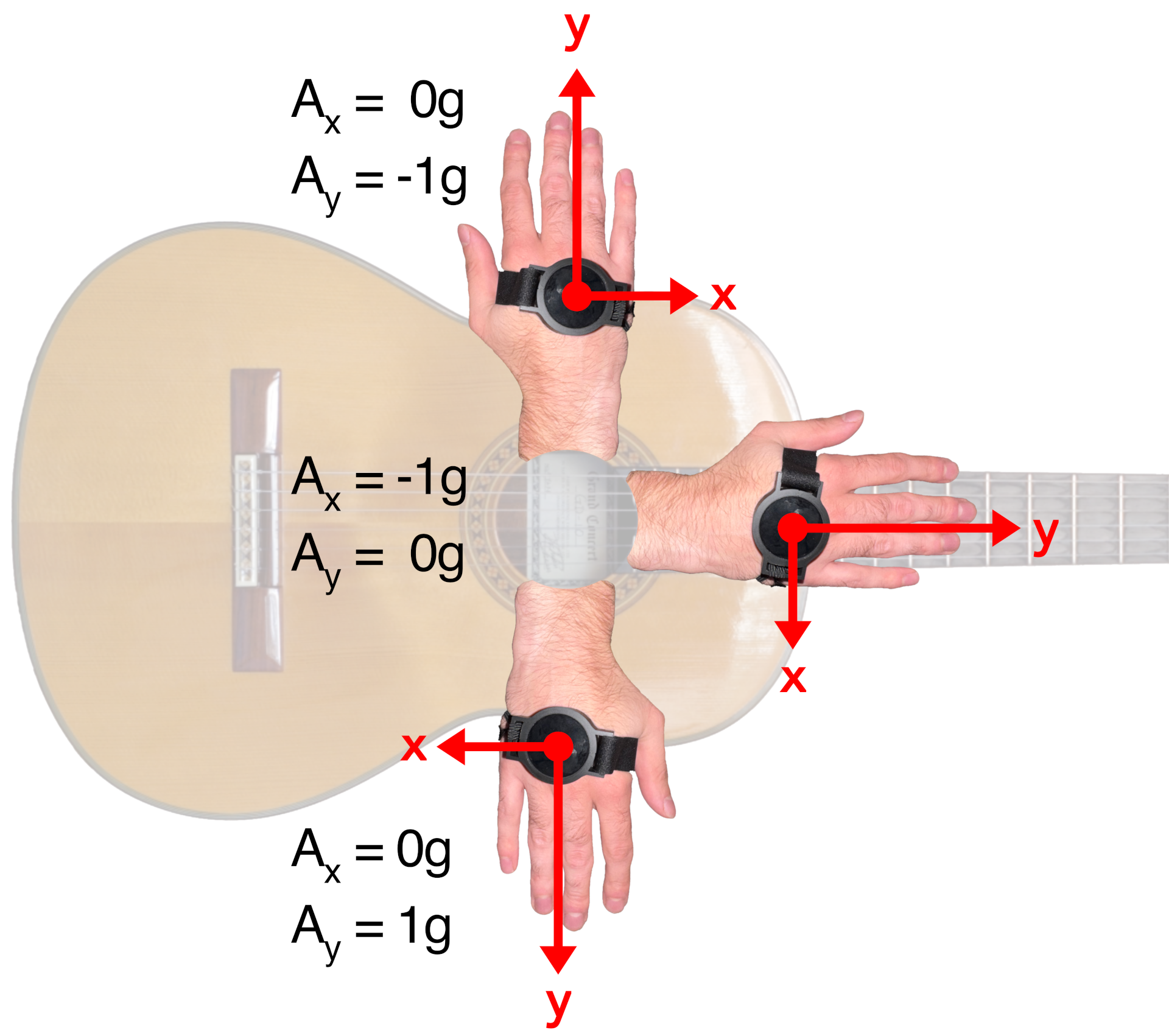}
  \caption{Motion model of the sensor}
  \label{fig:Bewegungsradius}
\end{figure}

Hand movement is described using a simplified model like in \cite{murgul2022multimodal}, in which the hand performs a semicircular motion around the elbow. The $x$-axis runs along the back of the hand, orthogonal to the fingers, while the $y$-axis is orthogonal to the $x$-direction, pointing towards the fingertips. The relevant acceleration components are gravitational acceleration $A_g$, centripetal acceleration $A_\text{centripetal}$, and tangential acceleration $A_\text{tangential}$ \cite{kleppner2014classical_mechanics}. The spatial orientations recorded by the sensor, along with the measured accelerations for different hand positions, are shown in Figure \ref{fig:Bewegungsradius}. The centripetal acceleration acts exclusively in the
$y$-direction, while the tangential acceleration occurs along the $x$-axis. Consequently, the acceleration $A_x$ in the $x$-direction and $A_y$ in the $y$-direction are given by
\begin{align}
  A_x & = A_{\text{tangential}} + A_g \cdot \cos(\phi)  \\
  A_y & = A_{\text{centripetal}} + A_g \cdot \sin(\phi)
\end{align}
where $\phi$ is the angle relative to the horizontal axis, ranging from $\SI{-90}{\degree}$ to $\SI{+90}{\degree}$. For slow, quasi-stationary hand movements, $A_x$ ranges from $-1\textsl{g}$ to $0\textsl{g}$, while $A_y$ takes values between $-1\textsl{g}$ and $1\textsl{g}$.
Due to the symmetry properties of the sine function, $A_x$ alone cannot determine the movement direction. However, by differentiating the acceleration in the y-direction, the movement direction can be inferred. A negative gradient corresponds to an upward motion, while a positive gradient corresponds to a downward motion.

In non-stationary cases, such as during strumming, both tangential and centripetal acceleration contribute to $A_x$ and $A_y$ respectively alongside the gravitational acceleration. The y-direction experiences an additional, constant centripetal acceleration. Because our method relies on acceleration derivatives, the constant centripetal acceleration can be ignored.

\subsection{Recording Process}
\label{sec:recording}

Table \ref{tab:recording_parameters} gives an overview of the playing instructions given to the guitarists.
To compile the datasets, $28$ different strumming patterns in 4/4 time signature based on \cite{Pattern1, Pattern2} are used, ranging from rhythmically simple to complex syncopated patterns. The patterns vary in parameters like tempo (60, 80, 100 BPM), chord progressions, playing style (plectrum, finger), and volume (soft, medium, loud). The variations were determined randomly based on a uniform distribution.

\input{tables/recording_parameters.tex}

The data collection was conducted with three guitarists, including a professional guitar teacher and two experienced amateur guitarists. The strumming patterns were played for $\SI{60}{\second}$ each to a metronome, following the predefined parameters. Simultaneously, audio recordings from the guitar pickup and acceleration data were captured. Synchronization of both audio signals was performed using cross-correlation. Additionally, the guitarists' playing was recorded using the microphone on an iPhone 15 Pro. The total recording duration amounts to $\SI{90}{\minute}$. Due to the lightweight design and the mounting position on the back of the hand, the guitar players' fingers were not constrained by the sensor during performance.

\subsection{Semi-Automatic Annotation}
\label{sec:annotation}

The annotation process involves identifying the onset times and strumming directions within the pickup recordings as well as the synchronization with the motion sensor signal. Instead of relying solely on automated onset detection, the process is optimized by incorporating prior knowledge from the recording plan, which includes tempo, rhythm patterns, chords, and strumming sequences. This structured information allows for a more robust prediction of expected onset times, reducing reliance on purely signal-based onset detection.
To determine actual onset times, spectral flux analysis \cite{bock2013specflux} is used to detect significant changes in the audio signal. However, since the guitarist does not necessarily start at the exact zero-second mark, a user-assisted graphical interface is employed to align the estimated onsets with the theoretical pattern. The process involves selecting the actual start time and iteratively adjusting until the detected onsets align with the expected timing based on the metronome.
Strumming direction is determined using acceleration data, which is synchronized with the audio signal. Since transmission latency and system delays introduce a time offset between the audio and acceleration data, manual adjustments are required. An interactive visualization displays both spectral flux and differentiated acceleration, allowing users to shift the acceleration data until the peaks of acceleration derivatives align with the detected onsets.

To assign strumming direction, peaks in the acceleration derivative corresponding to upward and downward hand movements are matched with detected onset peaks in spectral flux. If the acceleration derivative is positive at an onset time, it is labeled as an up strum; if negative, it is labeled as a down strum.
Next, we use the a priori information from the recording plan to automatically correct the annotations and add chord labels. Since we use a metronome, it can be assumed that the rhythmical pattern is played consistently enough to interpolate missed strumming events.
Finally, the annotated data undergoes manual validation and correction by a human annotator. The annotator visually inspects and adjusts the detected onsets and strumming directions using an interactive graphical interface.

%% file: tables/recording_parameters.tex
\begin{table}[h]
  \centering
  \begin{tabular}{ll}
    \toprule
    \textbf{Parameter} & \textbf{Values}                    \\
    \midrule
    Pattern            & 28 patterns                        \\
    Tempo              & 60 BPM, 80 BPM, 100 BPM            \\
    Movement           & little, normal, large              \\
    Volume             & quiet, medium, loud                \\
    Technique          & finger, pick                       \\
    Chords             & major and minor chord progressions \\
    \bottomrule
  \end{tabular}
  \caption{Results on microphone audio.}
  \label{tab:recording_parameters}
\end{table}

%% file: sections/003_synthesis.tex
\section{Guitar Strumming Synthesis}
\label{sec:synthesis}

To create a diverse and scalable dataset for training strumming transcription models, we introduce a novel strumming synthesis approach consisting of three stages: strumming tablature sampling, audio rendering, and audio augmentation. This method generates approximately $1000$ examples totaling $\SI{4}{\hour}$ of audio, which are randomly split into $\SI{90}{\percent}$ training, $\SI{5}{\percent}$ validation, and $\SI{5}{\percent}$ testing sets.

\subsection{Strumming Tablature Sampling}

\begin{figure*}[htb]
  \input{tikzpictures/strumming_sampling.tikz}
  \centering
  \caption{Flow chart of the strumming tablature sampling process.}
  \label{fig:strumming_sampling}
\end{figure*}
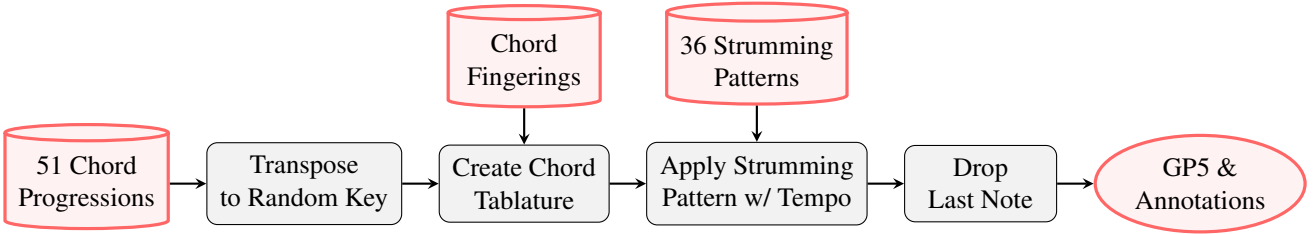

The first step involves generating synthetic strumming tablatures, as illustrated in Figure \ref{fig:strumming_sampling}. A database of $51$ chord progressions in functional notation and $36$ strumming patterns defined on a 16th-note grid serve as the foundation for generating variations. Each example is created by randomly selecting a chord progression, transposing it to a random key, and mapping each chord to a fingering from a lookup table. A random strumming pattern and tempo are then applied to create a complete tablature.
To introduce natural imperfections, the last note of a strumming chord is randomly dropped in $\SI{50}{\percent}$ of cases, simulating playing inconsistencies typical of amateur guitarists. The generated tablatures are stored in the GuitarPro\footnote{See https://www.guitar-pro.com for more information.} format, alongside a CSV annotation file containing timing, strumming action, and chord labels.

\subsection{Audio Rendering}

The synthesized tablatures are rendered into audio using DAWDreamer \cite{braun2021dawdreamer} and Ample Sound's virtual guitar instruments\footnote{Available at https://amplesound.net/en/index.asp.}, following a methodology similar to SynthTab \cite{zang2024synthtab}. Instead of converting tablatures to MIDI, we use \textit{.fxp} preset files to load the GuitarPro notation directly into the virtual instrument engine. This way, up and down stroke information can be input from the tablature.
To enhance realism, rendering parameters are randomized, including the blend between virtual microphones and the amount of fret noise introduced. The final output is saved as a $\SI{44.1}{\kilo\hertz}$ WAV file. Since the rendering process introduces an average $\SI{40}{\milli\second}$ latency, this delay is accounted for in the dataset annotations to maintain synchronization accuracy.

\subsection{Audio Augmentation}


To further improve realism and variability, a post-processing step applies a chain of effects using the Pedalboard library \cite{sobot2023pedalboard}. The augmentation pipeline introduces controlled distortions and environmental factors to better simulate real-world recordings.
The processing chain includes distortion, high- and low-pass filtering, and compression to mimic tonal variations across different recording conditions. To simulate room acoustics, a convolutional reverb effect is applied. Additional background noise layers, including ambient recordings (traffic, weather, and living room sounds) and white noise, are incorporated to model microphone imperfections and noisy environments. Finally, short bursts of fretting sounds and percussive noises, such as light tapping or clapping, are injected at random intervals to emulate natural guitar handling. The effect parameters, such as signal-to-noise ratio (SNR), filter cut-off frequencies, and dry/wet mix ratios, are randomized to ensure broad generalization.

%% file: tikzpictures/strumming_sampling.tikz
\begin{tikzpicture}[
    auto,
    node distance=0.48cm,
    roundnode/.append style={ellipse, draw=red!60, fill=red!5, very thick, minimum width = 2cm, minimum height = 1.2cm, align=center},
    process/.append style={rectangle, rounded corners, minimum width=2cm, minimum height=1cm, align=center, draw=black, fill=black!5, inner sep=5pt},
    database/.append style={cylinder, minimum width=2cm, minimum height=1cm, align=center, draw=red!60, fill=red!5, very thick, inner sep=5pt, shape border rotate=90, aspect=0.1},
    arrow/.style={thick,->,>=stealth}]

  \node (progression)[database] {51 Chord \\Progressions};
  \node (transpose) [process, right = of progression] {Transpose \\to Random Key};
  \node (tablature) [process, right = of transpose] {Create Chord\\ Tablature};
  \node (fingerings) [database, above = of tablature] {Chord \\Fingerings};
  \node (apply-pattern) [process, right = of tablature] {Apply Strumming \\Pattern w/ Tempo};
  \node (patterns) [database, above = of apply-pattern] {36 Strumming\\Patterns};
  \node (drop) [process, right = of apply-pattern] {Drop \\Last Note};
  \node (output) [roundnode, right = of drop] {GP5 \&\\Annotations};

  \draw [arrow] (progression) -- (transpose);
  \draw [arrow] (transpose) -- (tablature);
  \draw [arrow] (fingerings) -- (tablature);
  \draw [arrow] (tablature) -- (apply-pattern);
  \draw [arrow] (patterns) -- (apply-pattern);
  \draw [arrow] (apply-pattern) -- (drop);
  \draw [arrow] (drop) -- (output);

\end{tikzpicture}

%% file: sections/004_model.tex
\section{Model}
\label{sec:model}

Our model builds upon the Convolutional Recurrent Neural Network (CRNN) architecture proposed by Kong et al.\ \cite{kong2021high} for piano transcription. Unlike traditional classification-based approaches that estimate a discrete piano roll representation, this method employs a regression-based strategy to predict the time to the next onset or offset event. This design allows for more precise onset estimations beyond the limitations of fixed frame step sizes, while also increasing robustness against minor misalignments in onset label annotations during training.

\subsection{Pre-Processing}
\label{sec:preprocessing}
The input audio is resampled to $\SI{16}{\kilo\hertz}$ and segmented into overlapping $\SI{10}{\second}$ clips with a hop size of $\SI{1}{\second}$ to enhance data diversity.
Each segment is converted into a logarithmic Mel spectrogram, which serves as the input representation for the neural network. The spectrogram is computed using a window size of $2048$ samples and a hop size of $160$ samples, resulting in a time-frequency representation with $229$ frequency bins, starting at a minimum frequency of $\SI{30}{\Hz}$.
To improve generalization, random pitch shifts in the range $[-6, 6]$ semitones are applied during training, with chord labels transposed accordingly. The overlapping segmentation and augmentation ensure robust feature learning across diverse strumming patterns.

\subsection{Architecture}
\label{sec:architecture}


The model consists of two main components: a strumming onset regression network and a chord classification network. The input Mel spectrogram is first processed by a convolutional layer stack (Conv Stack) designed to capture time-frequency features.
The structure of the Conv Stack follows the design in \cite{kong2021high} and consists of four convolutional blocks. Each block contains two convolutional layers with identical kernel sizes, followed by a pooling operation that reduces the spectral dimension while preserving temporal information. After the final convolutional block, the extracted features are flattened for subsequent processing.
The flattened feature representation is passed through a fully connected (FC) layer before being fed into a bidirectional GRU (biGRU) layer with 256 units.
The output of the biGRU is then passed through another fully connected layer, which generates regression values for up strums and down strums.

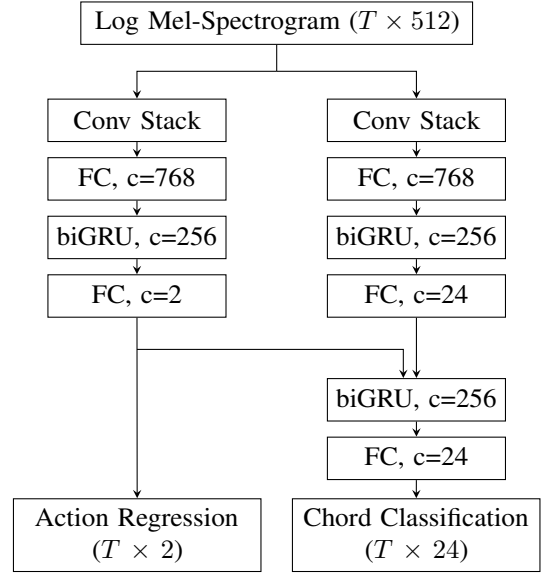
\begin{figure}
  \input{tikzpictures/model.tikz}
  \centering
  \caption{Joint strumming action detection and chord recognition network using logarithmic Mel spectrogram as input feature.}
  \label{fig:model}
\end{figure}

In parallel to the onset regression, a separate chord feature extraction stack processes the input spectrogram in a similar manner. Since chord labels are only available at strumming event times, the outputs of both networks are merged before passing through an additional biGRU and fully connected layer to produce final classification logits for 24 major and minor chord classes. Figure \ref{fig:model} provides an overview of the full model architecture.

\subsection{Regression Targets}
\label{sec:targets}

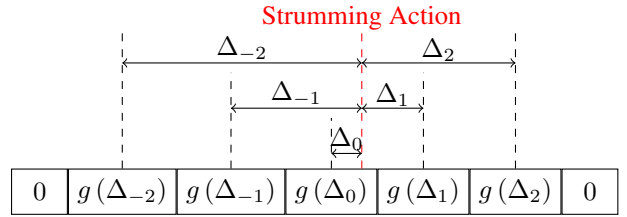
\begin{figure}[t]
  \input{tikzpictures/labels.tikz}
  \centering
  \caption{Structure of Strumming Action Onset Regression Labels.}
  \label{fig:regression_targets}
\end{figure}

Instead of relying on binary frame-based labels, a regression-based approach is used to determine strumming actions, as illustrated in Figure \ref{fig:regression_targets}.
The regression target function $g(\Delta_i) \in \left[0,1\right]$ encodes the time difference to the next strumming action onset $\Delta_i$, where $i$ is the index of a frame, using a triangular distribution. The target is defined as
\begin{equation} g(\Delta_i) =
  \begin{cases}
    1 - \frac{|\Delta_i|}{J\Delta}, & |i| \leq J     \\
    0,                              & |i| > J \quad,
  \end{cases}
\end{equation}
where $\Delta$ denotes the frame hop size and $J$ is a hyperparameter that controls the sharpness of the regression labels which is set to $5$ in our experiments.
The loss function consists of two components: one for strumming onset regression and another for chord classification. The strumming action regression loss $l_\text{action}$ is calculated from the regression output $R_\text{action}$ and the target $G_\text{action}$ by
\begin{equation}
  l_\text{action} = \sum_{t=1}^{T} \sum_{k=1}^{K} l_\text{bce} \left( G_\text{action}(t, k), R_\text{action}(t, k)\right) \quad ,
\end{equation}
where $l_\text{bce}$ represents the binary cross-entropy loss, $T$ is the number of time steps, and $K$ denotes the number of strumming action categories.
For chord classification, a similar loss function is used on the prediction outputs $P_\text{chord}$ and the targets $G_\text{chord}$:
\begin{equation}
  l_\text{chord} = \sum_{t=1}^{T} \sum_{c=1}^{C} l_\text{bce} \left( G_\text{chord}(t, c), P_\text{chord}(t, c)\right) \quad .
\end{equation}
where $C$ represents the number of possible chord labels. The total loss function used during training is simply the sum of both components:
\begin{equation}
  l = l_\text{action} + l_\text{chord} \quad .
\end{equation}


The model is trained using the AdamW optimizer \cite{loshchilov2017decoupled} with an initial learning rate of $10^{-4}$. The training process is run for $20,000$ steps with a batch size of $6$. On an NVIDIA Tesla V100 GPU, training takes approximately $\SI{2}{\hour}$.

%% file: tikzpictures/model.tikz
\tikzstyle{block} = [rectangle, draw, text width=6em, text centered, minimum height=1.6em]
\tikzstyle{block_in} = [rectangle, draw, text width=14em, text centered, minimum height=1.6em]
\tikzstyle{block_out} = [rectangle, draw, text width=8.6em, text centered, minimum height=1.6em]

\begin{tikzpicture}

  \node [block_in, rectangle] (mel) {Log Mel-Spectrogram ($T\times 512$)};

  \coordinate[below=1.0cm of mel] (dummy);

  \node [block, left=0.66cm of dummy] (conv1a) {Conv Stack};
  \node [block, below=0.2cm of conv1a] (fc1a) {FC, c=768};
  \node [block, below=0.2cm of fc1a] (gru1a) {biGRU, c=256};
  \node [block, below=0.2cm of gru1a] (fc2a) {FC, c=2};

  \node [block, right=0.66cm of dummy] (conv1b) {Conv Stack};
  \node [block, below=0.2cm of conv1b] (fc1b) {FC, c=768};
  \node [block, below=0.2cm of fc1b] (gru1b) {biGRU, c=256};
  \node [block, below=0.2cm of gru1b] (fc2b) {FC, c=24};

  \node [block, below=0.8cm of fc2b] (gru2) {biGRU, c=256};
  \node [block, below=0.2cm of gru2] (fc3) {FC, c=24};

  \coordinate[below=6.5cm of mel] (dummy2);

  \node [block_out, right=0.2cm of dummy2] (chord) {Chord Classification \\ ($T \times 24$)};
  \node [block_out, left=0.2cm of dummy2] (action) {Action Regression \\ ($T \times 2$)};

  \draw [-] (mel) -- ([yshift=+0.65cm]dummy);
  \draw [-stealth] ([yshift=+0.65cm]dummy) -| (conv1a);
  \draw [-stealth] ([yshift=+0.65cm]dummy) -| (conv1b);
  \draw [-stealth] (conv1a) -- (fc1a);
  \draw [-stealth] (fc1a) -- (gru1a);
  \draw [-stealth] (gru1a) -- (fc2a);
  \draw [-stealth] (conv1b) -- (fc1b);
  \draw [-stealth] (fc1b) -- (gru1b);
  \draw [-stealth] (gru1b) -- (fc2b);
  \draw [-stealth] (fc2b) -- (gru2);
  \draw [-stealth] (gru2) -- (fc3);
  \draw [-stealth] ([yshift=-0.42cm]fc2a.south) -| ([xshift=-0.4cm]gru2);

  \draw [-stealth] (fc2a) -- (action);
  \draw [-stealth] (fc3) -- (chord);

\end{tikzpicture}

%% file: tikzpictures/labels.tikz
\usetikzlibrary{calc}
\tikzstyle{block0} = [rectangle, draw, text width=0.5cm, text centered, minimum height=0.6cm]
\tikzstyle{block} = [rectangle, draw, minimum width=1.1cm, text centered, minimum height=0.6cm]

\begin{tikzpicture}

  \node [block] at (0, -3.12cm) (hrpt1) {$g\left(\Delta_{-2}\right)$};
  \node [block, right= 0cm of hrpt1] (hrpt2) {$g\left(\Delta_{-1}\right)$};
  \node [block, right= 0cm of hrpt2] (hrpt3) {$g\left(\Delta_{0}\right)$};
  \node [block, right= 0cm of hrpt3] (hrpt4) {$g\left(\Delta_{1}\right)$};
  \node [block, right= 0cm of hrpt4] (hrpt5) {$g\left(\Delta_{2}\right)$};
  \node [block0, right= 0cm of hrpt5] (hrpt6) {$0$};
  \node [block0, left= 0cm of hrpt1] (hrpt0) {$0$};

  \coordinate[right=0.4cm of hrpt3.north] (action);

  \node [color=red] at ([yshift=2.0cm]action) {Strumming Action};

  \draw[dashed, red](action)--([yshift=1.87cm]action);
  \draw[dashed](hrpt1.north)--([yshift=1.87cm]hrpt1.north);
  \draw[dashed](hrpt2.north)--([yshift=1.27cm]hrpt2.north);
  \draw[dashed](hrpt3.north)--([yshift=0.67cm]hrpt3.north);
  \draw[dashed](hrpt4.north)--([yshift=1.27cm]hrpt4.north);
  \draw[dashed](hrpt5.north)--([yshift=1.87cm]hrpt5.north);

  \draw[<->]([yshift=1.4cm]hrpt1.north)--([yshift=1.4cm]action);
  \draw[<->]([yshift=0.8cm]hrpt2.north)--([yshift=0.8cm]action);
  \draw[<->]([yshift=0.2cm]hrpt3.north)--([yshift=0.2cm]action);
  \draw[<->]([yshift=0.8cm]hrpt4.north)--([yshift=0.8cm]action);
  \draw[<->]([yshift=1.4cm]hrpt5.north)--([yshift=1.4cm]action);

  \node [above] at ($([yshift=1.3cm]hrpt1.north)!0.5!([yshift=1.3cm]action)$) {$\Delta_{-2}$};
  \node [above] at ($([yshift=0.7cm]hrpt2.north)!0.5!([yshift=0.7cm]action)$) {$\Delta_{-1}$};
  \node [above] at ($([yshift=0.12cm]hrpt3.north)!0.5!([yshift=0.12cm]action)$) {$\Delta_{0}$};
  \node [above] at ($([yshift=0.7cm]hrpt4.north)!0.5!([yshift=0.7cm]action)$) {$\Delta_{1}$};
  \node [above] at ($([yshift=1.3cm]hrpt5.north)!0.5!([yshift=1.3cm]action)$) {$\Delta_{2}$};

\end{tikzpicture}

%% file: sections/005_experiments.tex
\section{Experiments and Results}
\label{sec:experiments}

This section evaluates the performance of our proposed method for strumming onset detection, direction classification, and chord recognition. We begin by assessing the detection accuracy using guitar pickup signals, followed by an evaluation of real-world microphone recordings. Finally, we analyze the effectiveness of pitch shift augmentation and compare our chord recognition with existing approaches.

Model performance is measured using precision, recall, and F1-score for strumming detection. Specifically, we report these metrics for down strums ($\text{F1}_\text{down}$), up strums ($\text{F1}_\text{up}$), and strumming class agnostic ($\text{F1}_\text{any}$).
A $\SI{50}{ms}$ tolerance window is used, following the mir\_eval library \cite{raffel2014mireval}.

\subsection{Results on Guitar Pickup Signals}

In our first experiment, we explore the performance of our model directly on the guitar pickup signals. We use two of the guitarists we recorded to train our model and evaluate on the third guitarist.
We compare the detection quality of our trained model with common onset detection functions spectral flux \cite{bock2013specflux}, super flux \cite{bock2013specflux} and Complex Domain Onset Detection Function (CD-ODF) \cite{bello2004complexdomain} . For spectral flux and super flux, we use the implementation given in the librosa library \cite{mcfee2015librosa}.
The resulting precision, recall, and F1-score for any strumming direction are highlighted in Table \ref{tab:pickup_event_detection} for comparison.
Of the onset detection functions, the spectral flux offers the best detection results, directly followed by the CD-ODF. Compared with spectral flux and super flux, the CD-ODF offers a noticeably high recall. Therefore, it might be suitable for an active learning labeling scenario.
Our model outperforms the onset detection functions in all three precision, recall and F1-score. By achieving an F1-score of about $\SI{98}{\percent}$, the model is quite capable of reliably detecting the strumming actions in the pickup signal.

\input{tables/pickup_event_detection.tex}

By matching the detected strumming onsets with the movement data from the hand sensor, the strumming direction can also be determined. In Table \ref{tab:pickup_results}, we compare the results of the multimodal algorithmic approach with our CRNN model.
For all four approaches, the F1-score for down strums is higher than for up strums. Our CRNN model outperforms the algorithmic approaches for the down strum as well as the up strum class, whereby the increase is specifically noticable for up strum events.
Combining the CRNN detection with the acceleration-based classification leads to the overall best results. Therefore, the labeling could be automated quite efficiently by using a hybrid approach with the pickup audio signal to detect the events in the audio and the motion sensor data to get the strumming event class algorithmically.

\input{tables/pickup_result.tex}

\subsection{Results on Microphone Recordings}

\input{tables/mic_result.tex}

Next, we examine the action detection performance on the real-world microphone data. The real-world audio contains overall more noise, reverb and ambient sounds. The detection performance for different training dataset constellations (Synthetic (Sy), microphone exclusively (Ph), microphone and pickup (Ph + Pi), and all three datasets (Sy + Ph + Pi)) is compared in Table \ref{tab:mic_results}.
The $F1_\text{any}$ results for all datasets lie in a similar range. The synthetic dataset achieves about $\SI{5}{\percent}$ better results than when only using the comparably small training dataset of real-world phone recordings. When the pickup audio dataset is used in addition to the microphone recordings, we see a clear increase across all models. The increase is especially significant for up strums. In general, the real-world data performs significantly better than the synthetic dataset exclusively. Here, we see an increase of over $\SI{40}{\percent}$ compared to the synthetic dataset exclusively. Therefore, reliable onset detection itself can be trained from synthetic examples alone, but the classification of the strumming action profits from real-world audio.
The best overall results are obtained by combining the synthetic dataset with the microphone and pickup dataset. This indicates that increasing the real-world dataset in additional recording sessions might yield further improvements.
Interestingly, fine-tuning a checkpoint pretrained on synthetic data on the phone and pickup data leads to worse results than joining all three training datasets.

\subsection{Effect of Pitch Shift Augmentation}

\input{tables/augmentation_result.tex}

In the model pre-processing we perform data augmentation in the form of a random pitch shift before calculating the input spectrogram.
The effect of the pitch shift augmentation is studied using the training on the combined phone and pickup dataset. The results of this experiment are shown in Table \ref{tab:augmentation_results}. Applying a max pitch shift of 6 semitones leads to the best results. The F1-score for down strums increases by $\SI{10}{\percent}$ and up strums F1-score by $\SI{14}{\percent}$. While the pitch shift introduces more artifacts as the note shift increases, it also increases the diversity of chords used and therefore helps the model generalize.

\subsection{Chord Recognition}

\input{tables/chords_result.tex}

While the previous experiments only focused on the strumming action detection and classification, the chord recognition performance is quantified in this experiment and compared with a popular CNN-based \cite{korzeniowski2016feature} and a state-of-the art transformer model \cite{park2019bi}. We use the checkpoints provided by the authors. In contrast to the chord recognition task, where typically a musical piece is segmented into sections of a specific chord, we are interested in assigning a chord to a detected strumming event. Therefore, we use the ground truth strumming action times to determine a chord label. For the training of our own model, we use a maximum pitch shift of 6 semitones.
The resulting accuracy scores for the major-minor vocabulary are shown in Table \ref{tab:chords_result}.
The chord recognition transformer model and our model trained on the combined dataset achieve the best results of about $\SI{90}{\percent}$. The CNN-based chord tracking shows the weakest performance.
In contrast to the strumming action detection, our model trained on the synthetic dataset alone performs significantly better than with only the smaller real-world dataset. Training on all three datasets further increases the performance of our approach.

%% file: tables/pickup_event_detection.tex
\begin{table}[h]
  \centering
  \begin{tabular}{lccc}
    \toprule
    \textbf{Method}                       & $\text{F1}_\text{any}$          & $\text{P}_\text{any}$           & $\text{R}_\text{any}$           \\
    \midrule
    Spectral Flux \cite{bock2013specflux} & $\SI{79.49}{\percent}$          & $\SI{78.53}{\percent}$          & $\SI{81.86}{\percent}$          \\
    Super Flux \cite{bock2013specflux}    & $\SI{74.36}{\percent}$          & $\SI{77.04}{\percent}$          & $\SI{73.36}{\percent}$          \\
    CD-ODF \cite{bello2004complexdomain}  & $\SI{79.32}{\percent}$          & $\SI{68.50}{\percent}$          & $\SI{98.15}{\percent}$          \\
    \midrule
    \textbf{Ours}                         & \textbf{$\SI{97.60}{\percent}$} & \textbf{$\SI{96.54}{\percent}$} & \textbf{$\SI{98.73}{\percent}$} \\
    \bottomrule
  \end{tabular}
  \caption{Strumming detection results on pickup audio.}
  \label{tab:pickup_event_detection}
\end{table}

%% file: tables/pickup_result.tex
\begin{table}[h]
  \centering
  \begin{tabular}{lccc}
    \toprule
    \textbf{Methods}                      & $\text{F1}_\text{any}$          & $\text{F1}_\text{down}$         & $\text{F1}_\text{up}$           \\
    \midrule
    Spectral Flux \cite{bock2013specflux} & $\SI{79.49}{\percent}$          & $\SI{85.40}{\percent}$          & $\SI{68.60}{\percent}$          \\
    Super Flux \cite{bock2013specflux}    & $\SI{74.36}{\percent}$          & $\SI{84.40}{\percent}$          & $\SI{67.80}{\percent}$          \\
    CD-ODF \cite{bello2004complexdomain}  & $\SI{79.32}{\percent}$          & $\SI{82.20}{\percent}$          & $\SI{78.40}{\percent}$          \\
    \midrule
    \textbf{Ours}                         & \textbf{$\SI{97.60}{\percent}$} & $\SI{87.87}{\percent}$          & $\SI{84.90}{\percent}$          \\
    \textbf{Ours + Sensor}                & \textbf{$\SI{97.60}{\percent}$} & \textbf{$\SI{90.02}{\percent}$} & \textbf{$\SI{88.66}{\percent}$} \\
    \bottomrule
  \end{tabular}
  \caption{Strumming event detection results by class. The onset detection function results are paired with the hand movement signal in order to classify the events.}
  \label{tab:pickup_results}
\end{table}

%% file: tables/mic_result.tex
\begin{table*}[htb]
  \centering
  \begin{tabular}{l|ccc|ccc|ccc}
    \toprule
    \textbf{Training Data} & $F1_\text{any}$                 & $R_\text{any}$                  & $P_\text{any}$                  & $F1_\text{down}$                & $R_\text{down}$                 & $P_\text{down}$                 & $F1_\text{up}$                  & $R_\text{up}$                   & $P_\text{up}$                   \\
    \midrule
    Sy                     & $\SI{89.77}{\percent}$          & $\SI{89.47}{\percent}$          & $\SI{90.56}{\percent}$          & $\SI{73.92}{\percent}$          & $\SI{75.00}{\percent}$          & $\SI{74.04}{\percent}$          & $\SI{52.64}{\percent}$          & $\SI{56.99}{\percent}$          & $\SI{51.04}{\percent}$          \\
    Ph                     & $\SI{85.06}{\percent}$          & $\SI{84.11}{\percent}$          & $\SI{86.12}{\percent}$          & $\SI{79.90}{\percent}$          & $\SI{78.70}{\percent}$          & $\SI{81.42}{\percent}$          & $\SI{66.81}{\percent}$          & $\SI{67.52}{\percent}$          & $\SI{67.88}{\percent}$          \\
    Ph + Pi                & $\SI{89.45}{\percent}$          & $\SI{88.37}{\percent}$          & $\SI{90.64}{\percent}$          & $\SI{82.94}{\percent}$          & $\SI{83.72}{\percent}$          & $\SI{82.40}{\percent}$          & $\SI{75.10}{\percent}$          & $\SI{73.17}{\percent}$          & \textbf{$\SI{78.24}{\percent}$} \\
    \textbf{Sy + Ph + Pi}  & \textbf{$\SI{92.75}{\percent}$} & \textbf{$\SI{92.50}{\percent}$} & \textbf{$\SI{93.25}{\percent}$} & \textbf{$\SI{85.51}{\percent}$} & \textbf{$\SI{85.87}{\percent}$} & \textbf{$\SI{85.43}{\percent}$} & \textbf{$\SI{79.02}{\percent}$} & \textbf{$\SI{81.15}{\percent}$} & $\SI{77.80}{\percent}$          \\
    \bottomrule
  \end{tabular}
  \caption{Results on microphone audio trained on various combinations of the synthetic dataset (Sy), real-world pickup audio (Pi), and real-world microphone recordings (Ph).}
  \label{tab:mic_results}
\end{table*}

%% file: tables/augmentation_result.tex
\begin{table}[ht]
  \centering
  \begin{tabular}{llccc}
    \toprule
    \textbf{Max Pitch Shift}  & $F1_\text{any}$                 & $F1_\text{down}$                & $F1_\text{up}$                  \\
    \midrule
    None                      & $\SI{81.15}{\percent}$          & $\SI{71.04}{\percent}$          & $\SI{55.80}{\percent}$          \\
    $\pm3$ semitones          & $\SI{85.06}{\percent}$          & $\SI{79.10}{\percent}$          & $\SI{71.99}{\percent}$          \\
    \textbf{$\pm6$ semitones} & \textbf{$\SI{89.45}{\percent}$} & \textbf{$\SI{82.94}{\percent}$} & \textbf{$\SI{75.10}{\percent}$} \\
    $\pm12$ semitones         & $\SI{85.90}{\percent}$          & $\SI{80.89}{\percent}$          & $\SI{72.25}{\percent}$          \\
    \bottomrule
  \end{tabular}
  \caption{Effect of the max pitch shift parameter in the pre-processing step on the strumming detection performance.}
  \label{tab:augmentation_results}
\end{table}

%% file: tables/chords_result.tex
\begin{table}[ht]
  \centering
  \begin{tabular}{lccc}
    \toprule
    \textbf{Method (Dataset)}                                    & \textbf{Accuracy}               \\
    \midrule
    Deep Chroma Chord Recognition \cite{korzeniowski2016feature} & $\SI{80.37}{\percent}$          \\
    Chord Recognition BTC \cite{park2019bi}                      & $\SI{89.21}{\percent}$          \\
    \midrule
    Ours (Sy)                                                    & $\SI{87.84}{\percent}$          \\
    Ours (Ph + Pi)                                               & $\SI{81.52}{\percent}$          \\
    \textbf{Ours (Sy + Ph + Pi)}                                 & \textbf{$\SI{90.06}{\percent}$} \\
    \bottomrule
  \end{tabular}
  \caption{Results for chord recognition on the microphone audio of the real-world recordings.}
  \label{tab:chords_result}
\end{table}

%% file: sections/006_conclusion.tex
\section{Conclusion}
\label{sec:conclusion}
This study demonstrates the effectiveness of a CRNN-based model for the joint transcription of guitar strumming actions and chords. We introduced a novel approach to strumming synthesis, generating a large dataset of synthetic strumming examples. By extending an existing multimodal strumming transcription framework, we also collected 90 minutes of real-world guitar recordings, enhanced with semi-automatic annotations.
The combination of synthetic and real-world datasets allowed us to train a robust transcription model capable of accurately detecting strumming onsets, classifying strumming direction, and identifying chords from microphone audio.

Future work could extend this approach to cover a broader range of rhythmic patterns, including muted strumming events, which pose a challenge for motion-based annotation methods. Additionally, the chord vocabulary, currently limited to major and minor chords, could be expanded to include seventh chords, suspended chords, and other common chord voicings. These improvements would further enhance the versatility and real-world applicability of automatic strumming transcription models.